\documentclass[aps,prl, twocolumn, notitlepage]{revtex4-2} 

\usepackage{titlesec}

\usepackage{graphicx}
\usepackage{array}
\usepackage{float}
\usepackage{amsmath}
\usepackage{amssymb}
\usepackage{verbatim}
\usepackage{amsmath}
\usepackage{physics}
\usepackage{manfnt}
\usepackage{titlesec}
\usepackage[colorlinks=true]{hyperref}
\usepackage{siunitx}

\begin{document}
\title{Spin transport from order to disorder}
\author{Derek Reitz}
\author{Yaroslav Tserkovnyak}
\affiliation{Department of Physics and Astronomy and Bhaumik Institute for Theoretical Physics, University of California, Los Angeles, California 90095, USA}

\begin{abstract}
	Schwinger boson mean-field theory (SBMFT) is a non-perturbative approach which treats ordered and disordered phases of magnetic systems on equal footing. We leverage its versatility to evaluate the spin correlators which determine thermally-induced spin transport (the spin Seebeck effect) in Heisenberg ferromagnets (FMs) and antiferromagnets (AFs), at arbitrary temperatures. In SBMFT, the spin current, $J_s$, is made up of particle-hole-like excitations which carry integral spin angular momentum. Well below the ordering temperature, $J_s$ is dominated by a magnonic contribution, reproducing the behavior of a dilute-magnon gas. Near the transition temperature, an additional, paramagnetic-like contribution becomes significant. In the AF, the two contributions come with opposite signs, resulting in a signature, rapid inversion of the spin Seebeck coefficient as a function of temperature. Ultimately, at high temperatures, the low-field behavior of the paramagnetic SSE reduces to Curie-Weiss physics. Analysis based on our theory confirms that in recent experiments on gadolinium gallium garnet, the low-field spin Seebeck coefficient $\mathcal{S}(T) \propto \chi(T)$, the spin susceptibility, down to the Curie-Weiss temperature. At lower temperatures in the disordered phase, our theory shows a deviation of $\mathcal{S}(T)$ relative to $\chi(T)$ in both FMs and AFs, which increases with decreasing temperature and arises due to a paramagnetic liquid phase in our theory. These results demonstrate that the SSE can be a probe of the short-ranged magnetic correlations in disordered correlated spin systems and spin liquids.

\end{abstract}

\maketitle

\textit{Introduction.}| Most works in spintronics based on magnetic systems are asymptotic expansions or tailored phenomenological models which can be loosely divided into three categories: the strongly ordered regime that is handled by the Holstein-Primakoff approximation (HPA) and related treatements in 3D, the nonlinear-$\sigma$ model, or the Landau-Lifshitz-Gilbert phenomenology; the completely disordered paramagnetic Curie-Weiss regime; or criticality described by Landau theory. While the associated theories may work well in their respective small-parameter regimes, they fail outside of them. Moreover, phenomenology must be supported by an underlying fundamental description which contains the basic physical ingredients. The Schwinger boson transformation takes SU($\mathcal{N}$) generators to a product of $\mathcal{N}$ bosonic operators. The Hamiltonian is then decoupled by a Hubbard-Stratonovich transformation where the mean-field theory is the saddle point (SP), and the order $n$ fluctuations about the SP scale as $O(1/\mathcal{N}^n)$ \cite{arovas_1988, zhang_2021}. This approach, on the other hand, has no small or large parameter for fixed $\mathcal{N} \sim 1$, but still has the ability to qualitatively capture essential physics in regimes where we do not have an accurate theory. 

The spin Seebeck effect is generated by thermalized spin excitations and requires broken symmetry in spin space. Starting at $T \ll T_{C(N)}$ in ordered magnets, spin Seebeck coefficients theoretically \cite{Hoffman_2013, Uchida_2014, Rezende_2016, Okamoto_2016, Flebus_2019, Reitz_2020} and experimentally \cite{Prakash_2018, Wu_2016, Li2020} are generally expected to be enhanced by increasing temperature, while the opposite holds for paramagnets \cite{shiomi_2014, wu2015paramagnetic, liu_2018, Oyanagi_2019, Oyanagi_2021, Oyanagi_2023}, with the largest signals near the transition temperatures \cite{Uchida_2014, liu_2018, Li_2019, Yamamoto_2022}. These results suggest that the optimal regimes for thermoelectric applications may be distinct from the ones best described by HPA or the Curie-Weiss law, for example, which are designed to incorporate disorder or order, respectively, as minor corrections. In SBMFT, the FM, AF, and PM spin Seebeck coefficients reach their maxima around $T_{C(N)}$, where they reach the same order of magnitude when the Zeeman energy $\hbar \gamma B \approx J$, the exchange constant. While the SBMFT spin Seebeck coefficients in FMs and PMs have the same sign, in AFs the SSE inverts in sign slightly below $T_N$ due to the competition between antiferromagnetic and paramagnetic fluctuations.

The liquid-gas crossover in Heisenberg FMs and AFs appears as a continuous transition in SBMFT, and occurs at their Curie-Weiss temperatures $\Theta_{CW}$, with frustration parameter $f \equiv \abs{\Theta_{CW}}/T_{C(N)} \gtrsim 1$ in 3D. The liquid phase of the Heisenberg model in SBMFT is a simple setting for studying correlations effects in disordered spin systems, in 3D, as shown here, and also 2D \cite{Kim_2016, Samajdar_2019, Ghioldi_2018, Ghioldi_2022, bolsmann2023}. For example, by evaluating the spin correlators involved in thermally-induced spin transport across the paramagnetic phase, we show how spin Seebeck experiments can probe the properties of interacting spin liquids. SBMFT may play an important role for understanding spin transport measurements that can be used to manifest the magnetic properties of spin liquids \cite{Chatterjee_2015, Hirobe_2019}. This would complement indirect measurements such as the thermal conductivity and can support the limited information extracted from NMR and magnetic susceptibility measurements \cite{Savary_2016}. Along these lines, we introduce the parameter $p(T) \equiv \partial_B \mathcal{S} / \chi$, the ratio of the SSE to the spin susceptibility, which is $T$-independent when a magnet is completely disordered and becomes $T$-dependent when short-ranged spin correlations are significant to spin transport. $p(T)$ is then an indicator for spin correlations in the paramagnetic regime.

\textit{Mean-field theory.}| The Schwinger boson transformation replaces the spin operators by a product of bosonic creation and annihilation operators, $\mathcal{S}^+ = a^\dagger_{\uparrow}a_{\downarrow}$, $\mathcal{S}^- = a^\dagger_{\downarrow}a_{\uparrow}$, $S^z = \sum_\sigma \sigma a^\dagger_{\sigma}a_{\sigma}/2$, with the spin length fixed on each site by the constraint $S = \sum_\sigma a^\dagger_{\sigma}a_{\sigma}/2$. The SU(2)-preserving mean-field decomposition of the nearest-neighbor Heisenberg Hamilitonian on a bipartite lattice, written in terms of SBs $a_\sigma$ and $b_\sigma$ for sublattices $\mathcal{A}$ and $\mathcal{B}$, respectively, is
\begin{subequations}
	\begin{align}
		\label{Heis_Hamiltonian}
		H^{\mathrm{SU(2)}}_{\mathrm{mf}} = &-2J\sum_{\langle ij \rangle} \left[ \alpha F^\dagger_{ij} F - (1 - \alpha) A^\dagger_{ij}A  \right] + \mathrm{H.c.}\nonumber\\
		&-\mu_{\mathcal{A}} \sum_{i\in \mathcal{A}, \sigma} a^\dagger_{i\sigma} a_{i\sigma} - \mu_{\mathcal{B}} \sum_{i\in \mathcal{B}, \sigma} b^\dagger_{i\sigma} b_{i\sigma}.
	\end{align}
\end{subequations}
Here, summing over $\langle ij \rangle$ avoids double counting, $F_{ij} = \sum_\sigma a^\dagger_{i\sigma}b_{j\sigma}/2$ is a ``ferromagnetic'' contribution, and $A_{ij} = \sum_\sigma \sigma a_{i\sigma}b_{j\overline{\sigma}}/2$  is an ``antiferromagnetic'' contribution \cite{Messio_2013}.
These quartic terms are approximated in our MF decomposition by the product of a quadratic term and the mean fields $F = \langle F_{ij} \rangle$ and $A = \langle A_{ij} \rangle$, and in the same spirit the spin length constraints are implemented via two aggregate Lagrange multipliers $\mu_{\mathcal{A}(\mathcal{B})}$. This decomposition applies to isotropic lattice models where there is a single $F$ and single $A$ parameter. Note that while the exact constraint fixes the sum of the SB species' number operators on each site, $\mu_{\mathcal{A}(\mathcal{B})}$ instead fix the expectation value of this operator sum on each sublattice. $\alpha$ is a parameter that is free to vary in the exact Hamiltonian, but parameterizes separate mean-field Hamiltonians \cite{Messio_2013, zhang_2021}. To fix $\alpha$, we match the poles of the dynamic susceptibilities to the Holstein-Primakoff result at $T = 0$, giving the usual \cite{arovas_1988} $\alpha = 1$ for the FM and $\alpha = 0$ for the AF, and for simplicity fix these values for $\alpha$ at all $T$. In total, the bipartite FM (uniaxial AF below spin flop) has three mean-field parameters: $F$ ($A$), $\mu \equiv (\mu_{\mathcal{A}} + \mu_{\mathcal{B}})/2$, and $\delta \mu \equiv (\mu_{\mathcal{A}} - \mu_{\mathcal{B}})/2$. For the most general (Hartree-Fock-Bogoliubov) U(1)-preserving mean-field decomposition, see the Supplemental Material.

\begin{figure}[h]
	\includegraphics[width=0.48\textwidth]{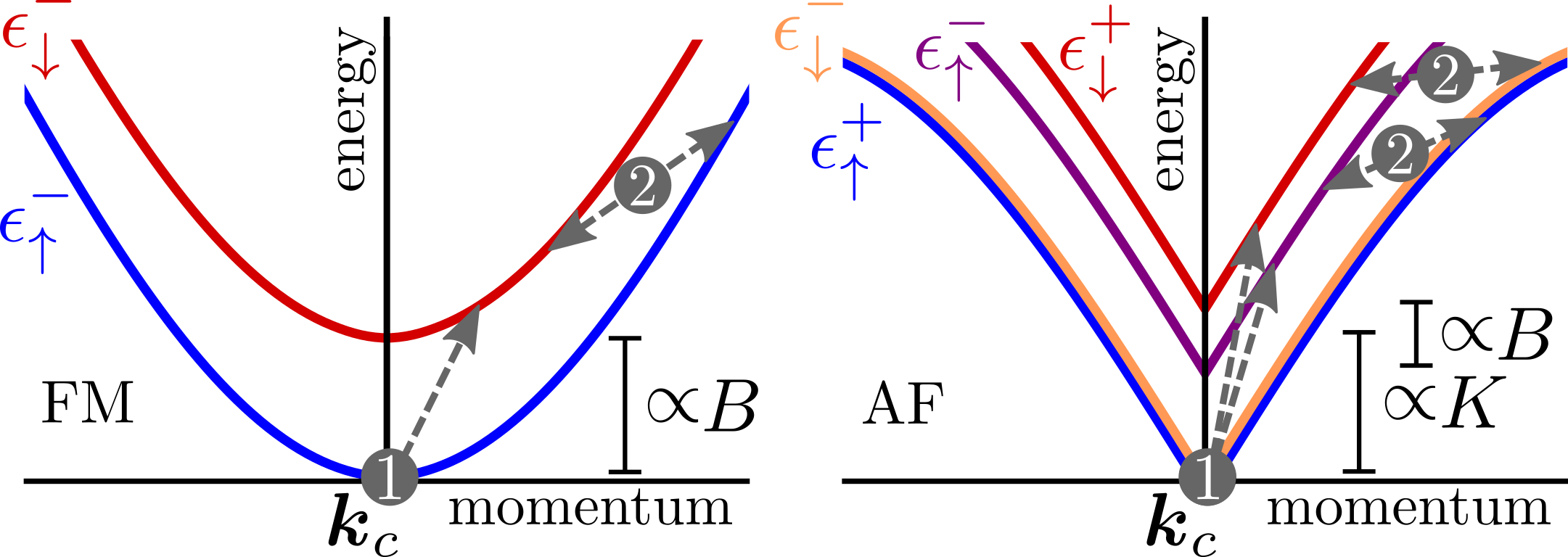}
	\caption{Schematic depiction of the magnonic (1) and paramagnetic-like (2) contributions to $J_s$. Each color specifies a combination of the bands' lower-indexed spin polarization and upper-indexed pseudospin. In SBMFT for FMs (AFs), at $T \leq T_{C(N)}$, Bose-Einstein condensation occurs at the lowest-energy modes with momentum $\boldsymbol{k_c}$. At $T > T_{C(N)}$ a self-consistent gap $-\mu$ opens up.}
	\label{fig:fig_depict}
\end{figure}

When $T \ll T_{C(N)}$, thermal equilibrium described by the Holstein-Primakoff picture is characterized by a dilute magnon gas with a single band for each sublattice \cite{Flebus_2019_prb}, which slightly depolarizes the spin ordering. In SBMFT, there are twice as many bands as in HPA, and each SB band carries half-integer spin. At a glance, the two pictures may seem irreconciliable. However, at $T_{C}$ in FMs the lowest-energy modes of one SB spin species (in the axially-symmetric case, for example) reach zero energy and form a Bose-Einstein condensate, resulting in long-ranged ordering along that species' spin polarization. At $T_N$ in AFs, long-ranged staggering ordering arises from condensation of one spin species on sublattice $\mathcal{A}$, and the opposite spin species on sublattice $\mathcal{B}$. Magnons in SBMFT are then spinful excitations associated with transitions from the condensates to the thermal cloud, as shown in Fig.~\ref{fig:fig_depict}. Thus, the SB bands on each sublattice which carry spin opposite to the local order mimick the magnon bands in Holstein-Primakoff. As we will see, these magnonic excitations will dominate spin transport at $T\ll T_{C(N)}$.

The SU(2)-preserving MFT yields a first-order Curie transition on cubic Bravais lattices, but is second-order on the diamond lattice, possibly due to its higher-order connectivity \cite{Tchernyshyov_2002}. The FM mean-field Hamiltonian plus applied field on the diamond lattice, setting $\delta \mu = 0$, after Fourier transforming and casting in terms of sublattice pseudospin, $\psi_{\boldsymbol{k}\sigma} = (a_{\boldsymbol{k}\sigma}, b_{\boldsymbol{k}\sigma})$, is
\begin{equation}
	H^{\mathrm{FM}}_{\mathrm{mf}} = \sum_{\boldsymbol{k}\sigma }\psi^\dagger_{\boldsymbol{k}\sigma} \left[ - (\mu + b \sigma/ 2) + \boldsymbol{\eta}_{\boldsymbol{k}} \cdot \boldsymbol{\tau} \right] \psi_{\boldsymbol{k}\sigma},
\end{equation}
where $b \equiv \hbar \gamma B$, $\boldsymbol{\eta}_{\boldsymbol{k}} = JF\left( -\Re \gamma_{\boldsymbol{k}}, \\ \Im \gamma_{\boldsymbol{k}}, \\ 0 \right)$, $\gamma_{\boldsymbol{k}} = Z^{-1} \sum_{\boldsymbol{\delta}} e^{i\boldsymbol{k}\cdot \boldsymbol{\delta}}$ is the structure factor, $\boldsymbol{\delta}$ is the vector between nearest neighbors on sublattice $\mathcal{A}$ to $\mathcal{B}$, and $\boldsymbol{\tau}$ is the vector of Pauli matrices. There are four bands with energies 
\begin{equation}
	\epsilon^\pm_{\boldsymbol{k}\sigma} = JZF(1\pm \abs{\gamma_{\boldsymbol{k}}}) - (\mu + b \sigma/ 2),
\end{equation}
where a factor of $JZF$ was absorbed into the definition of $\mu$. The eigenvectors are $v^{\pm}_{\boldsymbol{k}\sigma} =(1, \mp \abs{\gamma_{\boldsymbol{k}}} / \gamma_{\boldsymbol{k}}) /\sqrt{2}$. If $\mu$ reaches $-b/2$ the lowest energy branch, $\epsilon^-_{\boldsymbol{k}\uparrow}$, has zero-energy modes that condense, resulting in long-ranged spin ordering along the $+\hat{\boldsymbol{z}}$ axis in the language of SBs \cite{arovas_1988, Sarker_1989}. The lower-energy $\epsilon^-$ bands are shown in Fig.~\ref{fig:fig_depict}, and shown along with the high-energy $\epsilon^+$ bands in Supplemental Material Fig.~3. At arbitrary temperatures, the self-consistent mean-field equations for $F$ and $S$ give the solutions to $F(T)$ and either the condensate density $n_c(T)$ or $\mu(T)$ according to
\begin{equation}
	\label{mft}
	F = -(4N)^{-1} \sum_{\boldsymbol{k}\sigma \lambda} n^{\lambda}_{\boldsymbol{k}\sigma}\lambda \abs{\gamma_{\boldsymbol{k}}}, \;\;\; S = (4N)^{-1} \sum_{\boldsymbol{k} \sigma \lambda} n^{\lambda}_{\boldsymbol{k}\sigma},
\end{equation}
where $n^{\lambda}_{\boldsymbol{k}\sigma}$ is the Bose-Einstein distribution function for energy $\epsilon^{\lambda}_{\boldsymbol{k}\sigma}$, and $N$ is the number of sites per sublattice. In order to solve Eqs.~\eqref{mft} at $T<T_{C}$, the sums are converted to integrals with the contributions from the condensate density separated explicitly: for an arbitrary function $z$ and a single condensation point at momentum $\boldsymbol{k_c}$, $\sum_{\boldsymbol{k}} z_{\boldsymbol{k}}/N \approx z(\boldsymbol{k_c}) n_c + \mathcal{V} \int_{\mathrm{BZ}} d^3\boldsymbol{k} z(\boldsymbol{k}) / (2\pi)^3$, where $n_c \equiv N_c/N$ and $\mathcal{V}$ is the unit cell volume.

On the other hand, we find the N\'eel transition is second-order on all cubic Bravais lattices, so we take the simple cubic lattice for simplicity. The AF mean-field Hamiltonian with easy-axis anisotropy constant $K$ plus collinear applied field is
\begin{multline}
	H^{\mathrm{AF}}_{\mathrm{mf}} = \sum_{\boldsymbol{k}\sigma }\psi^\dagger_{\boldsymbol{k}\sigma} \left[\zeta_\sigma -(\delta \mu + b\sigma/2) \tau_z\right]\psi_{\boldsymbol{k}\sigma} +\\\sum_{\boldsymbol{k}\sigma} (i\sigma \psi^\intercal_{\boldsymbol{k}\sigma} \boldsymbol{\eta}_{\boldsymbol{k}} \cdot \boldsymbol{\tau} \psi_{-\boldsymbol{k}\overline{\sigma}}/2 +\mathrm{H.c.}),
\end{multline}
where we consider $b \ll \sqrt{JK}$, the spin-flop field; here $\zeta_\sigma = - \mu - K L^z \sigma/2$ for mean staggered spin polarization $L^z = (S^z_{\mathcal{A}} - S^z_{\mathcal{B}})/2$, $\boldsymbol{\eta}_{\boldsymbol{k}} = JA\left( \Im \gamma_{\boldsymbol{k}}, \\  \Re \gamma_{\boldsymbol{k}}, \\ 0 \right)$, and $\psi^\intercal$ is the vector transpose. Diagonalizing the Hamiltonian via a Bogoliubov transformation for each $\sigma$ yields four bands (see SM), we get energies
\begin{align}
	&\epsilon^{+}_{\boldsymbol{k}\sigma} = -\delta \mu - b\sigma/2 + \epsilon_{\boldsymbol{k}\sigma},\; \epsilon^{-}_{\boldsymbol{k}\sigma} = \delta \mu - b\sigma/2 + \epsilon_{\boldsymbol{k}\overline{\sigma}}, \\
	&\epsilon_{\boldsymbol{k}\sigma} \equiv \sqrt{\zeta_\sigma(2JZA+\zeta_\sigma) + (JZA)^2(1 - \gamma_{\boldsymbol{k}}^2)},\nonumber
\end{align}
where, like for the FM, we shifted $\mu$ by a factor of $JZA$, and $\overline{\sigma} = -\sigma$. Here, the ansatz $\delta \mu = -b/2$ was found by matching the field splitting of $\epsilon^{+}_{\boldsymbol{k}\downarrow}$ and $\epsilon^{-}_{\boldsymbol{k}\uparrow}$ to that of the usual AF magnon modes from HPA. This is a self-consistent solution for $T < T_N$, and then $\delta \mu = 0$ for $T \geq T_N$. Analogously to the FM, BEC occurs when the lowest-energy modes of $\epsilon^{+}_{\uparrow}$ and $\epsilon^{-}_{\downarrow}$ become gapless at $\mu = - K L^z /2 $, so that $\zeta_\sigma = KL^z(1-\sigma)/2$ \cite{Erlandsen_2020}, resulting in long-ranged staggered ordering. The modes are depicted in Fig.~\ref{fig:fig_depict}. The equations for $T<T_N$ are obtained by eliminating $n_c(T)$ to give two independent equations for $A(T)$ and $L^z(T)$, which in the limit $K \ll J$ (e.g., in Cr$_2$O$_3$, $K \approx 7 \cross 10^{-2}J$ \cite{Foner_1963}) are:
\begin{subequations}
	\label{af_mft}
	\begin{align}
		\label{A_mft} &A = S + C^A - (4N)^{-1} \sum_{\boldsymbol{k}\sigma } (n^+_{\boldsymbol{k}\sigma} + n^-_{\boldsymbol{k}\overline{\sigma}}) \sqrt{1-\gamma_{\boldsymbol{k}}^2}, \\
		\label{Lz_mft} &L^z = S - C^z - (2N)^{-1} \sum_{\boldsymbol{k}}(n^+_{\boldsymbol{k}\downarrow} + n^-_{\boldsymbol{k}\uparrow}) /\sqrt{1-\gamma_{\boldsymbol{k}}^2},
	\end{align}
\end{subequations}
where $C_A = 1/2 - (2N)^{-1}\sum_{\boldsymbol{k}} \sqrt{1-\gamma_{\boldsymbol{k}}^2} \approx 0.13$, $C_z = 1/2 - (N)^{-1}\sum_{\boldsymbol{k}} 1/\sqrt{1-\gamma_{\boldsymbol{k}}^2} \approx 0.25$, the contributions from the zero-energy modes vanish in Eq.~\eqref{A_mft}, and Eq.~\eqref{Lz_mft} only contains finite-energy modes. At $T>T_N$: $L^z = 0$ and $\mu(T)$ is no longer fixed so the mean-field equations are:
\begin{subequations}
	\begin{align}
		\label{A_eq_h} A &= (2N)^{-1} \sum_{\boldsymbol{k}\sigma} \left(n_{\boldsymbol{k}\sigma} + 1/2\right)\sqrt{ (-\mu + JZA)^2/\epsilon_{\boldsymbol{k}\sigma}^2 -1},\\
		\label{S_eq} S &= -1/2 +(2N)^{-1}\sum_{\boldsymbol{k}\sigma} \left(n_{\boldsymbol{k}\sigma} + 1/2\right) (-\mu + JZA) / \epsilon_{\boldsymbol{k}\sigma},
	\end{align}
\end{subequations}
where we took $n^+_{\boldsymbol{k}\sigma} \approx n^-_{\boldsymbol{k}\sigma} \equiv n_{\boldsymbol{k}\sigma}$ (valid when $K\ll J$).

\begin{figure}[h]
	\includegraphics[width=0.48\textwidth]{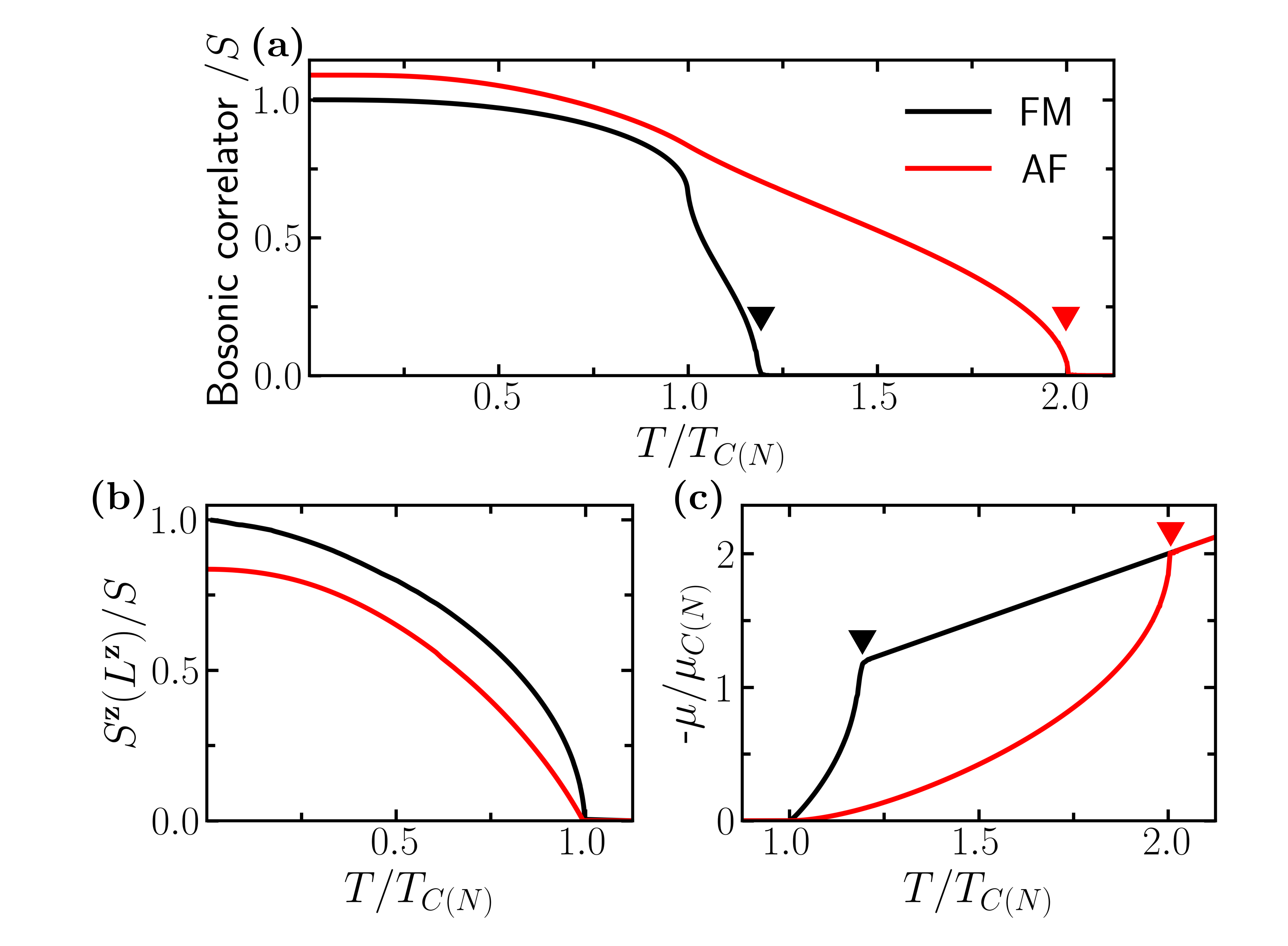}
	\caption{Mean-field solutions for the $S=1/2$ FM on the diamond lattice and the $S=3/2$ AF on the simple cubic lattice. For the FM (AF), (a) shows $F$ $(A)$, (b) shows $S^z$ ($L^z$) and (c) shows $-\mu$ in units of $\mu_{C(N)}= -T_{C(N)} \ln (1/S+1)$. Triangular markers denote the positions of the liquid-gas crossover.}
	\label{fig:fig_1}
\end{figure}

Finally, we compare the SBMFT magnonic excitations to the HPA dispersions in the strongly ordered phases. In the diamond-lattice FM, the lowest-energy modes of the $\epsilon^-_{\uparrow}$ band condense and the two $\epsilon^\pm_{\downarrow}$ bands match the magnon bands from HPA, which reproduces the usual Bloch $T^{3/2}$ law for demagnetization at $T\ll T_C$ \cite{Vargas_2020}. In the simple-cubic-lattice AF, the the lowest-energy modes of the $\epsilon^{+}_{\uparrow}$ and $\epsilon^{-}_{\downarrow}$ bands condense at $T_N$ forming staggered ordering while the $\epsilon^{+}_{\downarrow}$ and $\epsilon^{-}_{\uparrow}$ bands qualitatively match the magnon bands from HPA. They are $\epsilon^{+}_{\boldsymbol{k}\downarrow}, \epsilon^{-}_{\boldsymbol{k}\uparrow} = \pm b + \epsilon_{\boldsymbol{k}}$, where $\epsilon_{\boldsymbol{k}} = \sqrt{\epsilon_0^2 + (JZA)^2(1 - \gamma_{\boldsymbol{k}}^2)}$ with $\epsilon_0^2 = \epsilon_K(\epsilon_K + 2JZA)$ and $\epsilon_K = KL^z$. At $T \ll T_N$, the dispersive term $(JZA)^2(1 - \gamma_{\boldsymbol{k}}^2)$ with $A/S = 1+C_A/S$ differs by a constant factor from the HPA value, and the gap $\epsilon_0$ is proportional to $\epsilon_K = K(S-1/2)$ in HPA while it is $\epsilon_K = K(S-1/2+C_z)$ in SBMFT. The complete numerical solutions of the MFT for $B = 0$ with $S=1/2$ for the FM, where $n_c \propto S^z$, and $S=3/2$ for the AF, where $n_c \propto L^z$, are plotted in Fig.~\ref{fig:fig_1} ($T_C = 0.633J$ and $T_N = 5.12J$ in units where the Boltzmann constant $k_B = 1$).

\textit{Spin transport.}|
The net interfacial spin current between a magnetic insulator at $T_1$ and a metal at $T_2$ may be computed by treating the interfacial exchange Hamiltonian perturbatively with respect to the bulk. If we consider a ferromagnetic Bravais lattice with interfacial Hamiltonian in momentum space $H_{\mathrm{int}} = (V/N) \sum_{\boldsymbol{k},\boldsymbol{k'},q,q'} a^{\dagger}_{\boldsymbol{k} \uparrow} a_{\boldsymbol{k'} \downarrow} c^{\dagger}_{q \downarrow} c_{q'\uparrow} + \mathrm{H.c.}$, we get via FGR for the interfacial spin current density (in units of energy per area),
\begin{multline}
	\label{J_s_fm_cubic}
	J_s = \frac{g_{\uparrow \downarrow} }{2S N^2} \sum_{\boldsymbol{k}, \boldsymbol{k'}} \epsilon_{\boldsymbol{k}\boldsymbol{k'}\uparrow \downarrow} \cross \\ \left[ n_1(\epsilon_{\boldsymbol{k}\uparrow}) - n_1(\epsilon_{\boldsymbol{k'}\downarrow}) \right] \left[ n_1(\epsilon_{\boldsymbol{k}\boldsymbol{k'}\uparrow \downarrow} ) - n_2(\epsilon_{\boldsymbol{k}\boldsymbol{k'}\uparrow \downarrow}) \right],
\end{multline}
where $\epsilon_{\boldsymbol{k}\boldsymbol{k'}\uparrow \downarrow} \equiv \epsilon_{\boldsymbol{k}\uparrow} - \epsilon_{\boldsymbol{k'}\downarrow}$, and $g_{\uparrow \downarrow} \equiv 4\pi S D^2 V^2 / \mathcal{A}$ \cite{Bender_2015} is in units of inverse area where $D$ is the metal's density of states at the Fermi level in units of (energy$\cdot$volume$)^{-1}$ and $\mathcal{A}$ is the area per site of the interface. Eq.~\eqref{J_s_fm_cubic} shows that $J_s$ is made up of particle-hole like excitations which carry spin angular momentum. In the bipartite FM and AFs, the SBs on each sublattice split into mixtures of the two pseudospin SBs (for the full expressions for $J_s$ there, see the Supplemental Material). Finally, the spin Seebeck coefficient for $J_s(T_1, T_2)$ is defined as $\mathcal{S}(T) \equiv J_s(T+\delta T, T - \delta T)/\delta T$ in the limit $\delta T \ll T$ of linear response. 

In the ordered phases, the condensates grow macrospically large. In the thermodynamic limit, they must be separated from the integrals over the BZ. The contribution to the FM spin Seebeck coefficient on diamond due to the condensate density $n_c \propto S^z$ is
\begin{equation}
	\label{s_fm}
	\mathcal{S}^{\mathrm{FM}} = \frac{g_{\uparrow \downarrow} }{2s} S^z \int \frac{d^3\boldsymbol{k}}{(2\pi)^3 } \partial_T \left( \epsilon^+_{\boldsymbol{k}\downarrow} n^+_{\boldsymbol{k}\downarrow} + \epsilon^-_{\boldsymbol{k}\downarrow}n^-_{\boldsymbol{k}\downarrow} \right),
\end{equation}
where $s \equiv S/\mathcal{V}$, and $\epsilon^\pm_{\boldsymbol{k}\downarrow}$ are the magnon energies. For the AF, we consider an interface which is compensated in aggregate but is comprised of separate islands where the metal couples directly to either one of the two sublattices, and negligibly to the other \cite{Luo_2021, Flebus_2019}. In this scenario, the AF spin current is $J_s = J_s^{\mathcal{A}} + J_s^{\mathcal{B}}$, where $J_s^{\mathcal{A}}$ is generated by the coupling $H^{\mathcal{A}}_{\mathrm{int}} = (V/N) \sum_{\boldsymbol{k},\boldsymbol{k'},q,q'} a^{\dagger}_{\boldsymbol{k} \uparrow} a_{\boldsymbol{k'} \downarrow} c^{\dagger}_{q \downarrow} c_{q'\uparrow}+ \mathrm{H.c.}$ and $J_s^{\mathcal{B}}$ by $H^{\mathcal{B}}_{\mathrm{int}} = (V/N) \sum_{\boldsymbol{k},\boldsymbol{k'},q,q'}  b^{\dagger}_{\boldsymbol{k} \uparrow} b_{\boldsymbol{k'} \downarrow} c^{\dagger}_{q \downarrow} c_{q'\uparrow} + \mathrm{H.c.}$. The contribution to the AF spin Seebeck coefficient due to the condensate density $n_c \propto L^z$ is
\begin{equation}
	\label{s_af}
	\mathcal{S}^{\mathrm{AF}} = \frac{g_{\uparrow \downarrow} }{2s} L^z \int \frac{d^3\boldsymbol{k}}{(2\pi)^3} \frac{2JZA}{\epsilon^+_{\boldsymbol{k}\downarrow} + \epsilon^-_{\boldsymbol{k}\uparrow}} \partial_T \left( \epsilon^+_{\boldsymbol{k}\downarrow}n^+_{\boldsymbol{k}\downarrow} - \epsilon^-_{\boldsymbol{k}\uparrow}n^-_{\boldsymbol{k}\uparrow}\right).
\end{equation}
The AF SSE has contributions at the two magnon energies, $\epsilon^+_{\boldsymbol{k}\downarrow}$ and $\epsilon^-_{\boldsymbol{k}\uparrow}$, which come with opposite signs since they carry oppositely-oriented spin angular momentum. Eq.~\eqref{s_af} at $T\ll T_N$ reproduces the semiclassical N\'eel spin current derived in Ref.~\cite{Reitz_2020}.

\begin{figure}[h]
	\includegraphics[width=0.405\textwidth]{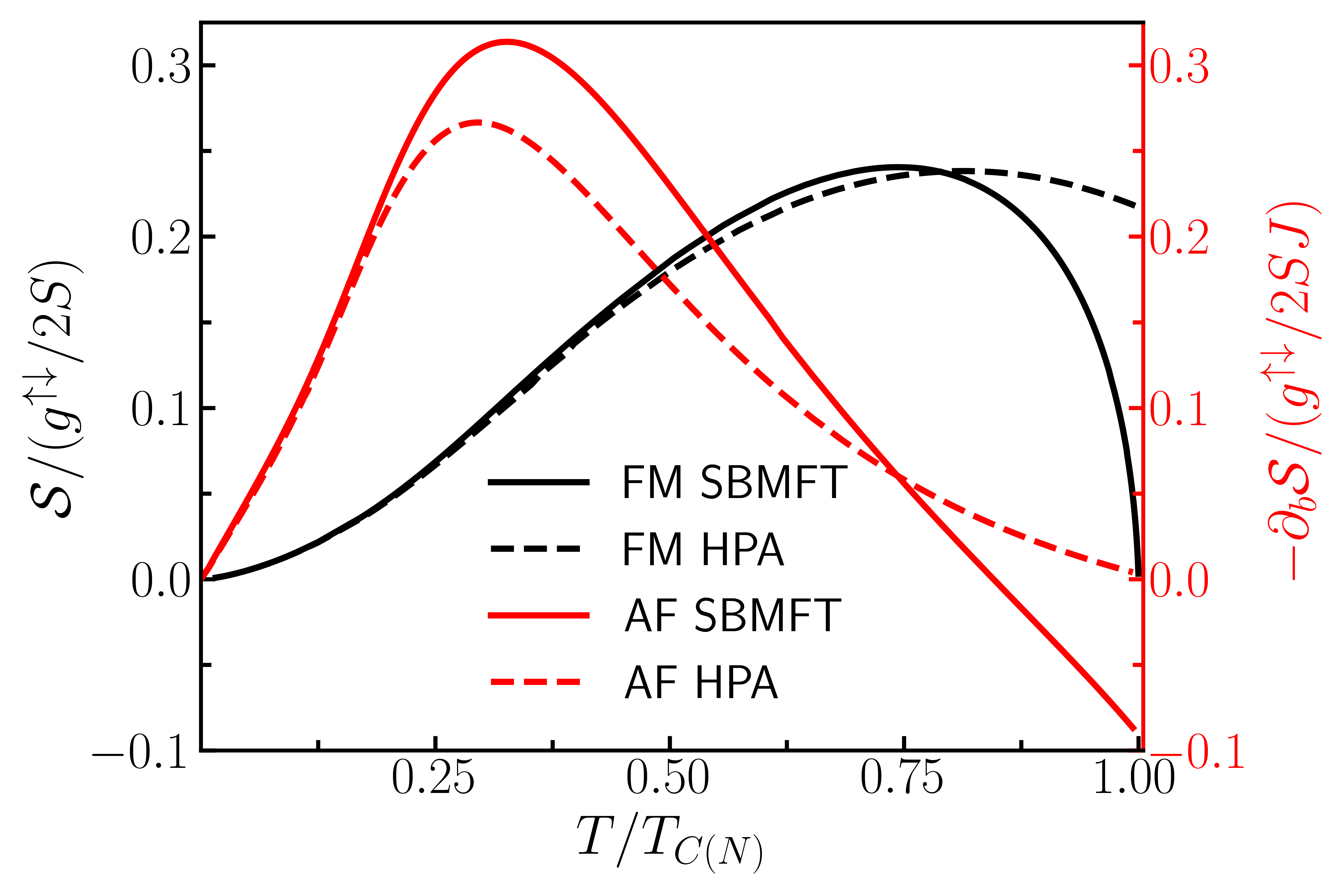}
	\caption{The spin Seebeck coefficients for the $S = 1/2$ FM on the diamond lattice and the negative field derivative $-\partial_b \mathcal{S}$ (with $b = \hbar \gamma B$ in units of $J$) for the $S = 3/2$ AF on the simple cubic lattice computed in the limit $B \rightarrow 0$ using SBMFT and HPA. }
	\label{fig:fig_2}
\end{figure}

At larger temperatures, $J_s$ also contains a contribution from scattering between bands in the thermal cloud, as shown in Fig.~\ref{fig:fig_depict}. This contribution is relatively smaller at $T \ll T_{C(N)}$ and becomes the paramagnetic spin current at $T > T_{C(N)}$. In order to carry out the two sets of integrals numerically in $\mathcal{S}^{\mathrm{PM}}$, we approximate the band structure with the low-energy, long-wavelength dispersion: $\epsilon^\pm_{\boldsymbol{k}\sigma} \approx JFk^2 - (\mu + b \sigma/ 2)$ for the FM and $\epsilon^\pm_{\boldsymbol{k}\sigma} \approx \pm (1 - \sigma)b/2 + \sqrt{\zeta_\sigma^2 - 2Z(JAk)^2}$ for the AF. The SBMFT spin Seebeck coefficients are compared to those computed in the same fashion using the Holstein-Primakoff transformation \cite{Bender_2015}, expanded to second order in the magnon over spin densities (defined as the Holstein-Primakoff approximation, HPA), and plotted as a function of temperature in Fig.~\ref{fig:fig_2}. 
\begin{figure}[h!]
	\includegraphics[width=0.35\textwidth]{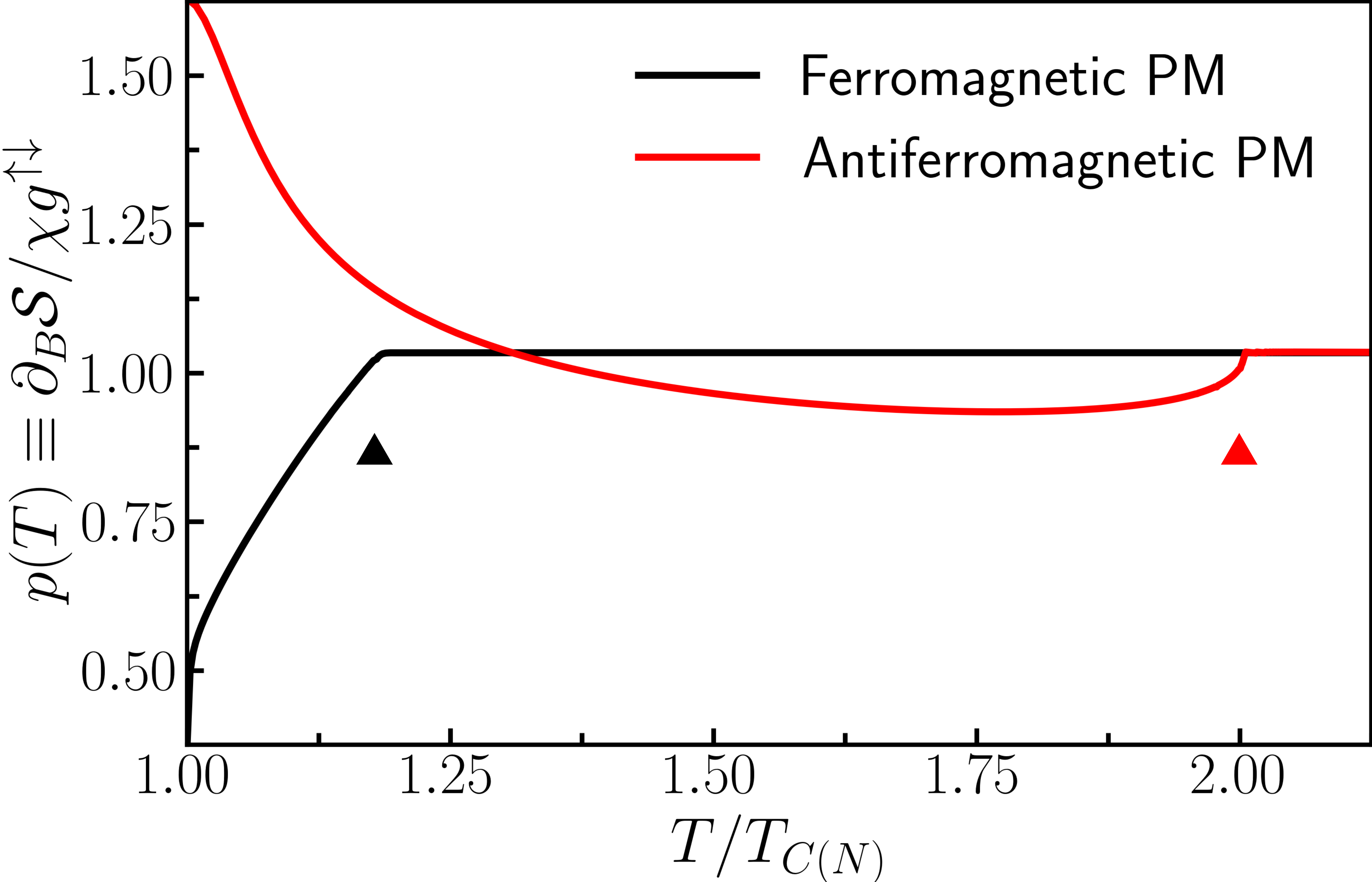}
	\caption{Field derivative of the paramagnetic SSE relative to the spin susceptibility in FMs and AFs. $\partial_B \mathcal{S}/g^{\uparrow \downarrow}$ begins to deviate from $\chi$ at the liquid-gas crossovers denoted by triangular markers.}
	\label{fig:fig_3}
\end{figure}

In strongly disordered spin systems, spin correlations decay on the scale of the lattice spacing. In SBMFT, this corresponds to $JF$, $JA \ll T$ and is described by the gaseous phase of the theory. In the gaseous phase at $b \ll T$, we get $\partial_B \mathcal{S}^{\mathrm{PM}} = \chi g^{\uparrow \downarrow}$ where $\chi \equiv \partial_B S^z / S$ is the normalized spin susceptibility. As $T$ decreases below $\Theta_{CW}$ in the SBMFT, this treatment has a continuous liquid-gas phase transition and spin correlations start to become significant. When $JF$ or $JA \sim T$, $\partial_B \mathcal{S}^{\mathrm{PM}}$ deviates from $\chi$. Based on this analysis of the Heisenberg model in SBMFT, we introduce a new frustration parameter $p(T) \equiv \partial_B \mathcal{S} / \chi$, whose temperature-dependence is an indicator for short-ranged spin correlations as shown in Fig.~\ref{fig:fig_3} (for comparison purposes, $\chi$ is also computed in the same fashion as $\mathcal{S}^{\mathrm{PM}}$ discussed above). 

\textit{Conclusion.}| Experimentally, extracting $p(T) \equiv \partial_B \mathcal{S} / \chi$ (Fig.~\ref{fig:fig_3}) is complicated since the measured spin Seebeck voltage, $V(B,T) = \mathcal{S}(B,T) f(T)$, contains additional temperature-dependent factors in $f(T)$, such as the interfacial thermal conductivity and metallic resistivity \cite{Reitz_2020, Oyanagi_2023}. However, we can analyze how the magnetic field profile, of the measured $V(B,T)$ and theoretical $\mathcal{S}(B,T)$, evolve with temperature. We illustrate this by comparing our theory for the SSE at $T \gg T_{C(N)}$ to experiments in gadolinium gallium garnet (GGG) \cite{wu2015paramagnetic, liu_2018}. We identify the field position of the peak in the SSE, at a given temperature, as a quantity which contains information about $\mathcal{S}(B,T)$, but is independent of $f(T)$. The peak data points are extracted from SSE field sweeps, and our theoretical values rely solely on the magnet's Curie-Weiss temperature. When we use an independently-measured value for $\Theta_{CW}$ from the static susceptibility in GGG \cite{PhysRevD.91.102004}, we find that our theory quantitatively reproduces the experimental SSE peak positions down to $T \geq 2\; \mathrm{K} \approx \Theta_{CW}$ (this is the lowest-temperature data currently available; for more details, see the Supplemental Material). At lower temperatures, a similar type of analysis could be used to investigate the emerging effects of short-ranged spin correlations in spin transport.

The sign change of the AF spin Seebeck coefficient as a function of temperature, below spin flop, at $T^* \approx 0.85T_N$ (Fig.~\ref{fig:fig_2}) is another feature which is insensitive to $f(T)$ because it is unlikely to change sign in the same region of $T$. The spin Seebeck coefficient in a Landau theory for the N\'eel transition has the paramagnetic sign \cite{yamamoto2019}, which is consistent with the SBMFT result in that the latter finds $T^*$ lies appreciably to the left of the transition temperature. While a bulk thermal gradient can drive an interfacial spin accumulation with the same sign as Eq.~\eqref{s_af} \cite{Rezende_2016}, this accumulation may be reduced and possibly invert in sign when Umklapp scattering becomes significant. It can reduce the magnon diffusion length and occurs when the temperature becomes comparable to the energy of magnons at the Brilluoin zone boundary. This occurs for the lower energy magnon branch before the higher energy branch, possibly leading to a lower value for $T^*$. To give a more quantitative estimate for $T^*$, a bulk spin transport theory for SBs must then be developed.

\textit{Acknowledgements.}|The work was supported by the U.S. Department of Energy, Office of Basic Energy Sciences under Award No. DE-SC0012190.
\bibliography{sbmft}
\end{document}


\title{Supplementary Material: Spin transport from order to disorder}
\author{Derek Reitz}
\author{Yaroslav Tserkovnyak}
\affiliation{Department of Physics and Astronomy and Bhaumik Institute for Theoretical Physics, University of California, Los Angeles, California 90095, USA}

\maketitle

\textit{General SB mean-field decomposition of Heisenberg Hamiltonian.}| In magnets, U(1)-symmetry is broken spontaneously in the ordered phase. In the SU(2)-preserving SBMFT decomposition discussed in the main text (which is the standard used throughout the field of SB research including in ordered magnets, e.g. \cite{arovas_1988, okamoto_2016, Samajdar_2019, Erlandsen_2020, Vargas_2020, Zhang_2022}), the symmetry in the ordered phase most be broken explicitly by an infinitesimal applied field. Additionally, disordered magnets in the presence of applied fields are known to obey the Curie-Weiss law, which arises from a U(1)-symmetric effective field felt by a spin due to the weak polarization of neighboring spins. Thus, when performing our SB mean-field decomposition of the Heisenberg Hamiltonian, we should attempt to consider all terms which respect U(1) symmetry. The most general (Hartree-Fock-Bogoliubov) U(1)-preserving mean-field decomposition of the nearest-neighbor Heisenberg Hamilitonian on a bipartite lattice, written in terms of SBs $a_\sigma$ and $b_\sigma$ for sublattices $\mathcal{A}$ and $\mathcal{B}$, respectively, is
\begin{subequations}
	\begin{align}
		\label{Heis_Hamiltonian}
		&H = -J \sum_{\langle ij \rangle} \mathbf{S}_i \cdot \mathbf{S}_j = -2J\sum_{\langle ij \rangle} \left[ \alpha F^\dagger_{ij}F_{ij} - (1 - \alpha) A^\dagger_{ij}A_{ij} \right] - \sum_{i\sigma}\lambda_i(a^\dagger_{i\sigma} a_{i\sigma} - 2S),\\
		&H_{\mathrm{mf}} = H_{\mathrm{mf}}^{\rm SU(2)} + H_{\mathrm{mf}}^{\rm U(1)},\\
		&H_{\mathrm{mf}}^{\rm SU(2)} = -2J\sum_{\langle ij \rangle} \left[ \alpha F^\dagger_{ij} F - (1 - \alpha) A^\dagger_{ij}A + \mathrm{H.c.} \right] -\mu_{\mathcal{A}} \sum_{i\in \mathcal{A}, \sigma} a^\dagger_{i\sigma} a_{i\sigma} - \mu_{\mathcal{B}} \sum_{i\in \mathcal{B}, \sigma} b^\dagger_{i\sigma} b_{i\sigma},\\
		&H_{\mathrm{mf}}^{\rm U(1)} = -JZ\left( \sum_{i \in \mathcal{A}} S_{\mathcal{B}}^z S^z_{i} + \sum_{i \in \mathcal{B}} S_{\mathcal{A}}^z S^z_{i} \right) -2J\sum_{\langle ij \rangle, \sigma} \left[ \alpha C^\dagger_{ij, \sigma}C_{\overline{\sigma}} - (1 - \alpha)D^\dagger_{ij, \sigma}D_{\overline{\sigma}} + \mathrm{H.c.} \right].
	\end{align}
\end{subequations}
Here, summing over $\langle ij \rangle$ avoids double counting, $F_{ij} = \sum_\sigma a^\dagger_{i\sigma}b_{j\sigma}/2$, $A_{ij} = \sum_\sigma \sigma a_{i\sigma}b_{j\overline{\sigma}}/2$, $C_{ij, \sigma} = a_{i\sigma}b_{j\overline{\sigma}}$ and $D_{ij, \sigma} = a^\dagger_{i\sigma}b_{j\sigma}$. The mean-fields are $F = \langle F_{ij} \rangle$ and similarly for $A, C, D$, $S^z_{\mathcal{A}(\mathcal{B})}$ is the mean $z$-component of spin on the $\mathcal{A} (\mathcal{B})$ sublattice, and $J Z S_{\mathcal{A}(\mathcal{B})}^z $ is the effective field which yields the Curie-Weiss law for $Z$ nearest neighbors. 

Note that solutions to the SU(2) problem are consistent with $C = D = 0$, but the effective field $J Z S_{\mathcal{A} (\mathcal{B})}^z$ is generally finite in the ordered phase, or when SU(2) symmetry is broken by an applied field in the disordered phase. Therefore, we do not have a reason \textit{a priori} to discard it. Starting from $T=\infty$, we find a gaseous paramagnetic phase with $F = A = 0$ that survives at finite temperatures down to the liquid-gas transition. In the gaseous phase, the susceptibility follows the Curie-Weiss law. This U(1) term also provides spontaneous symmetry breaking at the Curie/N\'eel transition. However, it introduces a magnon gap in the ordered phase, where we expect gapless Goldstone modes in the Heisenberg FM and AF, and results in first-order Curie and N\'eel transitions. There is an indication that the issues may be resolved, for example, by taking the full U(1) decomposition for a given $\alpha$ in Eq.\eqref{Heis_Hamiltonian}. When we take $\alpha = 1$ at $T\ll T_C$ and consider both $F\neq 0$ and $C_{\sigma} \neq 0$, we find a low-$T$ solution which restores the Bloch law. However, since this significantly increases the complexity of the MF equations and is not within the scope of this work, we do not consider finite $C$ and $D$ further and instead call for future investigations into a more complete SBMFT in the ordered phase. The effect of treating $\alpha$ as a separate mean-field parameter is also an open question, though Ref.~\cite{Ghioldi_2022} used it as a fitting parameter to reproduce data on the dynamical spin structure factor of a $S=1/2$ triangular lattice antiferromagnet, obtaining $\alpha = 0.436$.
 
\begin{figure}[h!]
	\includegraphics[width=0.5\textwidth]{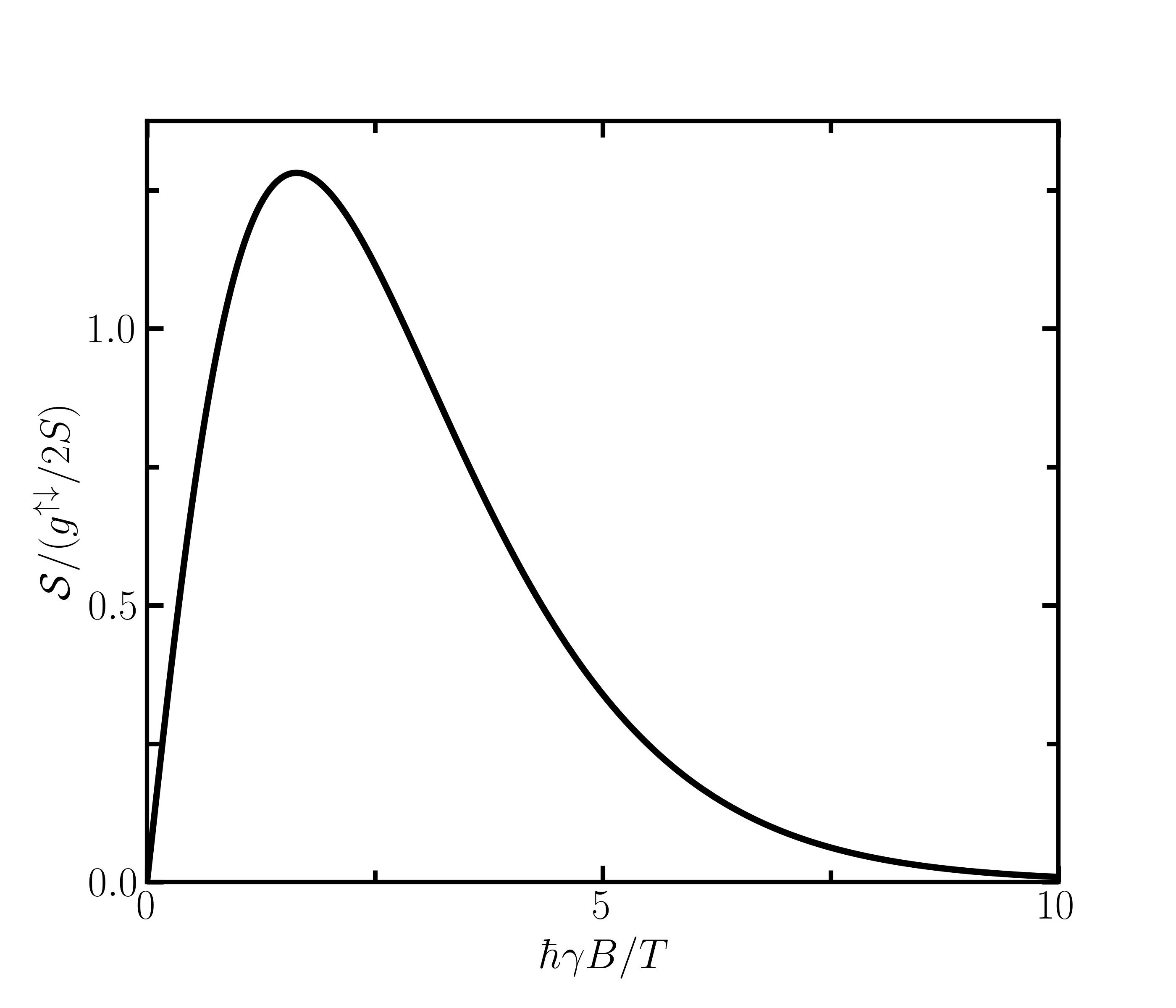}
	\caption{The gaseous paramagnetic spin Seebeck coefficient, Eq.~\eqref{s_pm}, with $\Theta_{CW} = 0$ as a function of $b/T$.}
	\label{fig:fig_1}
\end{figure}

\textit{U(1) symmetric mean-field solution at $T \gg T_{C(N)}$.}| In this section we consider the U(1) symmetric MF Hamiltonian with applied field $B\hat{\boldsymbol{z}}$ in the gaseous ($F = A = 0$) paramagnetic phase and self-consistently take the fields $C = D = 0$. The FM and AF mean-field equations have the same form: there are two energy levels which are split by magnetic field, and two mean-field parameters $\mu$ and $S^z$. The mean-field equations are
\begin{subequations}
	\label{mft}
	\begin{align}
		\label{S_mft} &S = \sum_{\sigma} n_{\sigma}/2, \\
		\label{Sz_mft} &S^z = \sum_{\sigma} \sigma n_{\sigma}/2,
	\end{align}
\end{subequations}
where $\epsilon_{\sigma} = - (\mu + h \sigma/ 2)$ with $h \equiv \hbar \gamma B + J Z \langle S^z \rangle$, which includes the U(1) term from the mean-field decomposition of the Heisenberg Hamiltonian. At $h \ll T$ the solution (setting $ K_B = 1$) is $\mu = -T \ln(1/S + 1) - h/2$ and $S^z/S = \chi B$ where $\chi = \mathcal{C} / (T - \Theta_{CW})$ is the spin-length normalized susceptibility with $\mathcal{C} = (S+1)\hbar\gamma/2$ and $\Theta_{CW} = J Z \mathcal{C}$, where the Curie constant $\mathcal{C}$ is a constant factor of $3/2$ larger than the usual one. In the gaseous phase the spin Seebeck coefficient in linear response is
\begin{equation}
	\label{s_pm}
	\mathcal{S}^{\mathrm{PM}} = -\frac{g_{\uparrow \downarrow} }{2S} \left[ n(\epsilon_{\uparrow}) - n(\epsilon_{\downarrow}) \right] \epsilon_{\uparrow \downarrow} \partial_T n(\epsilon_{\uparrow \downarrow} ) = -\frac{g_{\uparrow \downarrow} S^z}{S} h \partial_T n(h),
\end{equation}
where $\epsilon_{\uparrow \downarrow} \equiv \epsilon_{\uparrow} - \epsilon_{\downarrow}$. The full numerical results with $J=0$ are plotted in Fig.\ref{fig:fig_1}; when $J > 0$ or $J < 0$ the curve is shifted left and right, respectively, such that the peak in $\mathcal{S}^{\mathrm{PM}}(B)$ occurs at $b \approx T - \Theta_{CW}$.

\begin{figure}[h!]
	\includegraphics[width=0.8\textwidth]{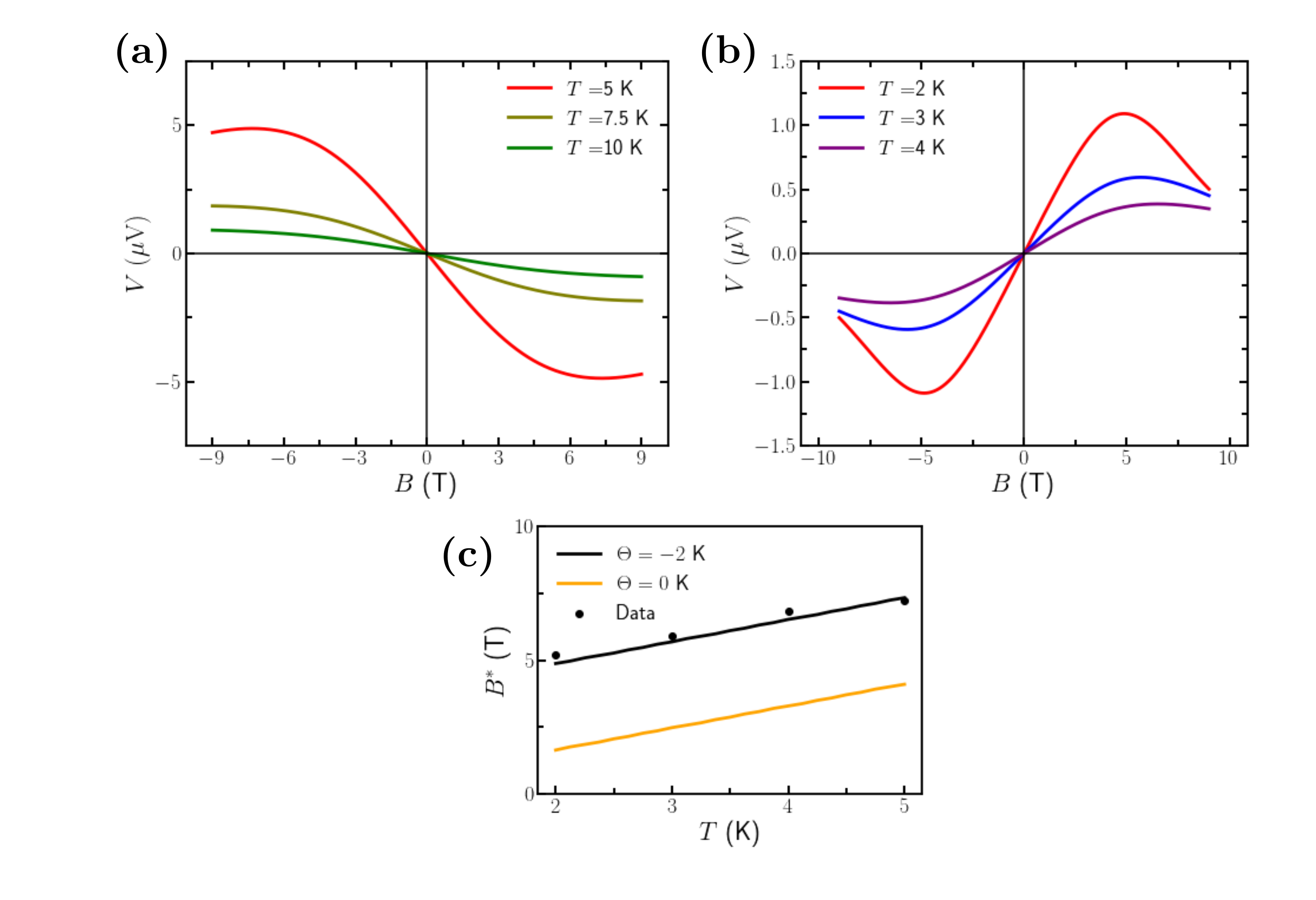}
	\caption{Applied field-dependent spin Seebeck voltage data from Refs.~\cite{wu2015paramagnetic} in (a) and \cite{liu_2018} in (b), is reproduced by the gaseous phase spin Seebeck coefficient, Eq.~\eqref{s_pm}, times a temperature dependent factor taken from the data where $\Theta_{CW} = -2$ K is used. In (c) the magnetic field where the SSE is maximized is plotted ($T = 2, 3, 4$ K from Ref. \cite{liu_2018} and $T = 5$ K from Ref. \cite{wu2015paramagnetic}) which depends only on the spin Seebeck coefficient. Panel (c) gives $\Theta_{CW} = -2$ K (antiferromagnetic $J$) as the best fit, which agrees with independent magnetic susceptibility measurements on GGG. }
	\label{fig:fig_2}
\end{figure}

\textit{Comparison to SSE data on GGG.}| In this section we reproduce the temperature evolution of the non-monotonic field dependent SSE observed in GGG by S. Wu, C. Liu, et al. \cite{wu2015paramagnetic, liu_2018}. The measured spin Seebeck voltage $V(B,T) = \mathcal{S}(B,T) f(T)$ contains additional temperature-dependent factors parameterized by $f(T)$ \cite{Reitz_2020}. In Ref.~ \cite{wu2015paramagnetic}, they found $\partial_B V |_{B\rightarrow 0} \propto T^{-3.38}$, and using the three full field sweeps at $T = 2,3,4$ K from Ref.~\cite{liu_2018} we obtain $\partial_B V |_{B\rightarrow 0} \propto T^{-2.45}$ for that device. When the voltage slope follows a power law in $T$, we can fit to $f(T)$ at $T \gg T_{C(N)}$ using our theory for $\mathcal{S}(B,T)$.  After also absorbing $g_{\uparrow \downarrow}/2S$ into $f(T)$, the normalized spin Seebeck voltage in the gaseous phase of the SBMFT is $\partial_B V / f = \chi$, where $J$ is the only undetermined parameter in $\chi$. The magnetic susceptibility of GGG is well known and from the theory we have the relation $\Theta_{CW}/\mathcal{C} = JZ$ where the Curie-Weiss temperature of $\Theta_{CW} = 2.10$ K is taken from Ref.~\cite{PhysRevD.91.102004}. This allows us to plot the theoretical curves shown in Fig.~\ref{fig:fig_2}a,b. We can also extract the field $B^*(T)$ where $V$ is maximized as a function of field, a quantity that is independent of $f(T)$, as a quantitative comparison to our theory shown in Fig.~\ref{fig:fig_2}c. From our theory's point of view, application of the gaseous phase SSE results to this data is consistent until near $T = \abs{\Theta_{CW}}$, where the theory's liquid-gas phase transition occurs. At $T$ somewhat lower than $\abs{\Theta_{CW}}$ there may be enhancements to spin transport resulting in $\partial_B V / f \neq \chi$ as discussed in the main text.

\textit{FM mean field equations with $n_c$.}| At $T \leq T_C$, the equations for $F$ and $S$ contain Bose-Einstein distribution numbers of zero-energy modes which condense into the density $n_c$. To obtain solutions numerically, we instead work with mean-field equations where the $n_c$ contributions have been eliminated. The quantities $F-S$ and $S^z - S$ have this property:
\begin{subequations}
	\label{mft}
	\begin{align}
		\label{fm_F_mft} &F = S - \frac{1}{4N} \sum_{\boldsymbol{k}\sigma \lambda} n^{\lambda}_{\boldsymbol{k}\sigma}(1 + \lambda \abs{\gamma_{\boldsymbol{k}}}), \\
		\label{fm_Sz_mft} &S^z = S - \frac{1}{2N} \sum_{\boldsymbol{k} \lambda} n^{\lambda}_{\boldsymbol{k}\downarrow},
	\end{align}
\end{subequations}
where the contribution from the zero-energy modes now vanishes in Eq.~\eqref{fm_F_mft}, and Eq.~\eqref{fm_Sz_mft} only contains finite-energy modes; thus, they can now be converted into integrals directly. Note that the two equations for $F$ and $S^z$ are also decoupled. Finally, at $B = 0$, $S^z = n_c/4$. The mean field solutions for $F(T)$ and $\mu(T)$ at $T > T_C$ cannot be decoupled and are obtained by solving the equations in the main text. 

\textit{Diagonalization of AF Hamiltonian.}| The Bogoliubov transformation (very similar to, c.f. Ref.~\cite{Rezende_2016}, the usual Bogoliubov transformation for AF magnons in the Holstein-Primakoff approximation; specifically, when $\sigma = \downarrow$, our transformation matches the structure of the magnonic one) that diagonalizes the AF Heisenberg Hamiltonian after a SU(2)-symmetric mean-field decomposition, plus uniaxial anisotropy and collinear applied field below spin flop, is
\begin{subequations}
	\begin{align}
		(a_{\boldsymbol{k}\sigma}, b_{-\boldsymbol{k}\overline{\sigma}}) &= u_{\boldsymbol{k}\sigma} \Psi_{\boldsymbol{k}\sigma} + \sigma v_{\boldsymbol{k}\sigma} \tau_x \Psi^\dagger_{\boldsymbol{k}\sigma},\\
		u_{\boldsymbol{k}\sigma}^2 &= \frac{\zeta_\sigma + JZA}{\epsilon^+_{\boldsymbol{k}\sigma}+\epsilon^-_{\boldsymbol{k}\overline{\sigma}}} + 1/2, \; v_{\boldsymbol{k}\sigma}^2 = \frac{\zeta_\sigma + JZA}{\epsilon^+_{\boldsymbol{k}\sigma}+\epsilon^-_{\boldsymbol{k}\overline{\sigma}}} - 1/2,
	\end{align}
\end{subequations}
where $\tau_x$ is the pauli matrix, $\zeta_\sigma \equiv -\mu - K L^z \sigma/2$, $\Psi_{\boldsymbol{k}\sigma} \equiv (\alpha_{\boldsymbol{k}\sigma}, \beta_{-\boldsymbol{k}\sigma})$ and $\alpha, \beta$ are the Bogoliubon field operators whose energies are $\epsilon^+_{\boldsymbol{k}\sigma}, \epsilon^-_{\boldsymbol{k}\sigma}$, respectively. Here, the anisotropy term in the Hamiltonian was treated in the same fashion as the Heisenberg Hamiltonian -- a quartic term decomposed, such that SU(2)-symmetry is preserved, here as products of quadratic operators times the mean field $L^z$. The mean-field equations are
\begin{subequations}
	\begin{align}
		\label{A_eq} A &= \frac{1}{2N} \sum_{\boldsymbol{k}\sigma} \gamma_{\boldsymbol{k}} (n^+_{\boldsymbol{k}\sigma} + n^-_{\boldsymbol{k}\overline{\sigma}} + 1)u_{\boldsymbol{k}\sigma}v_{\boldsymbol{k}\sigma},\\
		\label{S_eq} S &= \frac{1}{4N}\sum_{\boldsymbol{k}\sigma} \left[ (n^+_{\boldsymbol{k}\sigma}+n^-_{\boldsymbol{k}\overline{\sigma}})(u^2_{\boldsymbol{k}\sigma}+v^2_{\boldsymbol{k}\sigma}) + 4 v^2_{\boldsymbol{k}\sigma}\right],
	\end{align}
\end{subequations}
where Eq.~\eqref{S_eq} is the mean-field constraint. Other quantities of interest are
\begin{subequations}
	\begin{align}
		\label{Lz_eq} L^z &=  \frac{1}{4N}\sum_{\boldsymbol{k}\sigma }\sigma \left[ (n^+_{\boldsymbol{k}\sigma} + n^-_{\boldsymbol{k}\overline{\sigma}})(u^2_{\boldsymbol{k}\sigma}+v^2_{\boldsymbol{k}\sigma}) + 4 v^2_{\boldsymbol{k}\sigma}\right], \\
		\label{Sz_eq} S^z &=  \frac{1}{4N}\sum_{\boldsymbol{k}\sigma} \sigma (n^+_{\boldsymbol{k}\sigma} - n^-_{\boldsymbol{k}\overline{\sigma}}).
	\end{align}
\end{subequations}
When $T \leq T_N$, $\mu = -K L^z /2 $ and the condensate density is separated from the sums. Note that the condensate contributions, which go as $n_c (u^2_{\boldsymbol{k_c}\uparrow}+v^2_{\boldsymbol{k_c}\uparrow})$, where $\boldsymbol{k_c}$ are the condensation points in the BZ, appear to diverge since $(u^2_{\boldsymbol{k_c}\uparrow}+v^2_{\boldsymbol{k_c}\uparrow}) \propto 1/(\epsilon^+_{\boldsymbol{k_c}\uparrow}+\epsilon^-_{\boldsymbol{k_c}\uparrow})$. However, the energies only go to zero in the thermodynamic limit and so they cannot be taken to zero without simultaneously taking the system size to infinity \cite{Erlandsen_2020}. Inserting the Bogoliubov coherence factors, separating the condensate density from the sums, and using Eq.~\eqref{S_eq} to eliminate it from Eqs.~\eqref{A_eq} and \eqref{Lz_eq}, we get
\begin{subequations}
	\label{af_mft}
	\begin{align}
		\label{A_mft} &A = S + 1/2 + \frac{1}{4N}\sum_{\boldsymbol{k}\sigma} (n^+_{\boldsymbol{k}\sigma} + n^-_{\boldsymbol{k}\overline{\sigma}} + 1) \left[\gamma_{\boldsymbol{k}}\sqrt{\left(\frac{2\zeta_\sigma + 2JZA}{\epsilon^+_{\boldsymbol{k}\sigma}+\epsilon^-_{\boldsymbol{k}\overline{\sigma}}} \right)^2 - 1} -  \frac{2\zeta_\sigma + 2JZA}{\epsilon^+_{\boldsymbol{k}\sigma}+\epsilon^-_{\boldsymbol{k}\overline{\sigma}}}\right], \\
		\label{Lz_mft} &L^z = S + 1/2 - \sum_{\boldsymbol{k}} (n^+_{\boldsymbol{k}\downarrow} + n^-_{\boldsymbol{k}\uparrow} + 1)\frac{\zeta_\sigma + JZA}{\epsilon^+_{\boldsymbol{k}\downarrow}+\epsilon^-_{\boldsymbol{k}\uparrow}},
	\end{align}
\end{subequations}
where $\delta_{i,j}$ is the Kronecker delta. Eqs.~\eqref{A_mft} and \eqref{Lz_mft} no longer contain contributions from the condensate terms, so the sums can now be converted to integrals in order to solve numerically. When $T\geq T_N$, we have $L^z = 0$ and $\mu(T)$ is no longer fixed. If $ K \ll J$ so that anisotropy can be neglected, then $\epsilon^+_{\boldsymbol{k}\sigma} \approx \epsilon^-_{\boldsymbol{k}\sigma} \equiv E_{\boldsymbol{k}\sigma}$, $u_{\boldsymbol{k}\uparrow} \approx u_{\boldsymbol{k}\downarrow}$ and $v_{\boldsymbol{k}\uparrow} \approx v_{\boldsymbol{k}\downarrow}$, which reproduces the equations in the main text. For $S^z$ we get
\begin{equation}
	\label{Sz_eq_h} S^z =   \frac{1}{2N}\sum_{\boldsymbol{k}\sigma} \sigma n_{\boldsymbol{k}\sigma}.
\end{equation}
In the limit $ K \ll J$, these reproduce the equations in the main text.

\begin{figure}[H]
	\centering
	\includegraphics[width=0.3\textwidth]{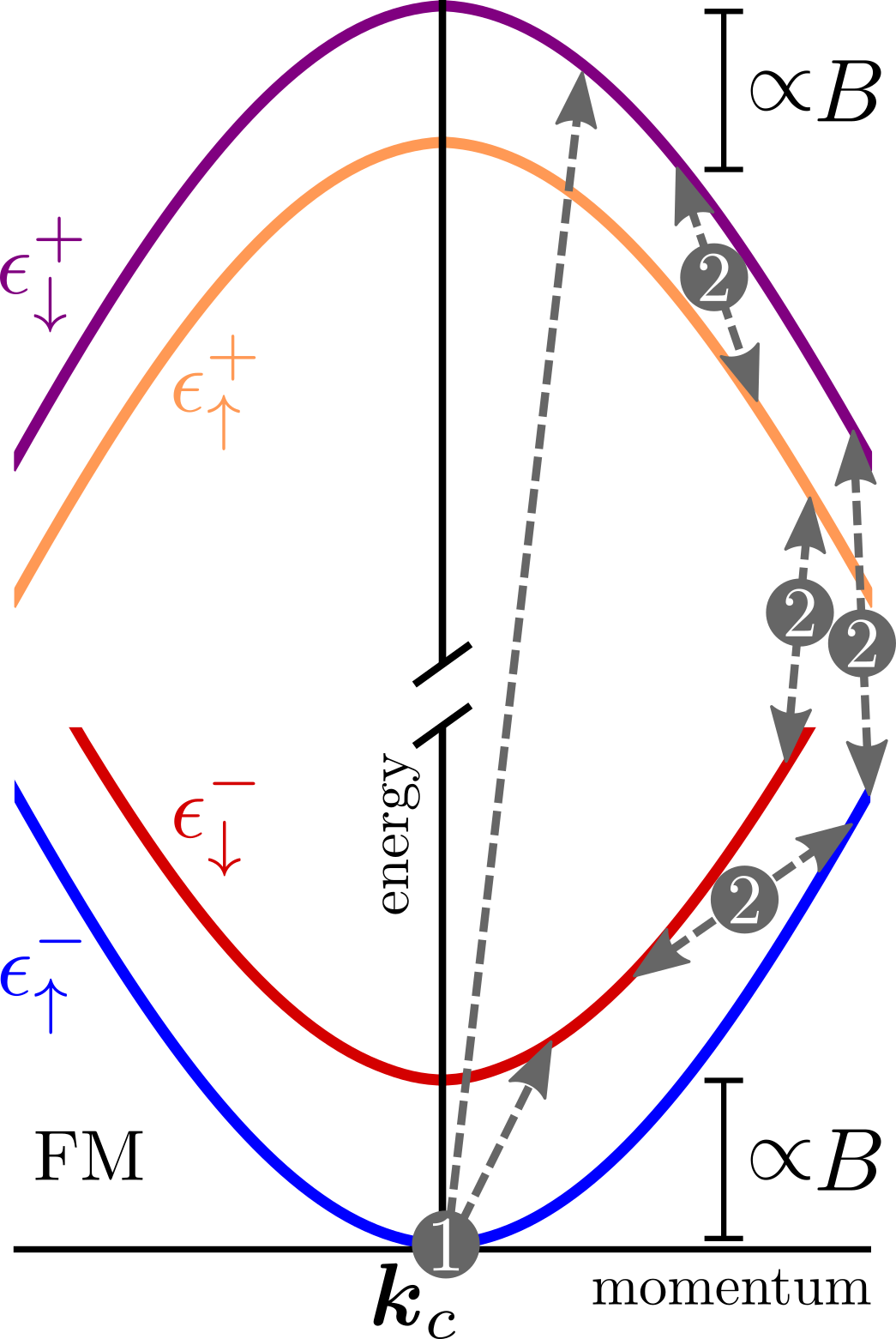}
	\caption{Schematic depiction of the scattering processes which make up $J_s$ in the FM SBMFT on diamond. At $T>T_C$, all bands are gapped by $\mu$. As $T\rightarrow T_C$, $\mu \rightarrow 0$, and the lowest energy mode of the $\epsilon^-_{\uparrow}$ band condenses, as shown here. Near the condensation point $\boldsymbol{k_c}$, the $+$ bands are maximal. Processes labeled by (1) are magnonic excitations, which dominate $J_s$ at $T\ll T_C$; while (2) result in a paramagnetic-like contribution to $J_s$, which vanishes linearly as $B\rightarrow 0$. }
	\label{fig:fig_3}
\end{figure}

\textit{Full expressions for the SBMFT spin currents.}| In this section we compute $J_s = J_s^{\mathcal{A}} + J_s^{\mathcal{B}}$, where $J_s^{\mathcal{A}}$ is generated by the coupling $H^{\mathcal{A}}_{\mathrm{int}} = (V/N) \sum_{\boldsymbol{k},\boldsymbol{k'},q,q'} a^{\dagger}_{\boldsymbol{k} \uparrow} a_{\boldsymbol{k'} \downarrow} c^{\dagger}_{q \downarrow} c_{q'\uparrow}+ \mathrm{H.c.}$, and $J_s^{\mathcal{B}}$ by $H^{\mathcal{B}}_{\mathrm{int}} = (V/N) \sum_{\boldsymbol{k},\boldsymbol{k'},q,q'}  b^{\dagger}_{\boldsymbol{k} \uparrow} b_{\boldsymbol{k'} \downarrow} c^{\dagger}_{q \downarrow} c_{q'\uparrow} + \mathrm{H.c.}$. For the FM on the diamond lattice $J_s^{\mathcal{A}} = J_s^{\mathcal{B}} = J_s/2$, and we get
\begin{equation}
	\label{J_s}
	J_s = \frac{g_{\uparrow \downarrow} }{2S N^2} \sum_{\nu,\nu',\boldsymbol{k}, \boldsymbol{k'}}  \epsilon^{\nu \nu'}_{\boldsymbol{k}\boldsymbol{k'}\uparrow \downarrow} \left[ n_1(\epsilon^{\nu}_{\boldsymbol{k}\uparrow}) - n_1(\epsilon^{\nu'}_{\boldsymbol{k'}\downarrow}) \right] \left[ n_1(\epsilon^{\nu \nu'}_{\boldsymbol{k}\boldsymbol{k'}\uparrow \downarrow} ) - n_2(\epsilon^{\nu \nu'}_{\boldsymbol{k}\boldsymbol{k'}\uparrow \downarrow}) \right],
\end{equation}
where $\epsilon^{\nu \nu'}_{\boldsymbol{k}\boldsymbol{k'}\uparrow \downarrow} \equiv \epsilon^{\nu}_{\boldsymbol{k}\uparrow} - \epsilon^{\nu'}_{\boldsymbol{k'}\downarrow}$, and the sums run over the pseudospin indices $\nu, \nu' = +,-$. In the FM on diamond, $J_s$ contains both intraband and interband scattering, as shown in Fig.~\ref{fig:fig_3}. Note that our SBMFT constraints fix the expectation value of the number of $\uparrow$ and $\downarrow$ SBs on each sublattice. However, the diagonal basis of the Hamiltonian is in terms of spinon operators $\alpha_\sigma$ and $\beta_\sigma$, whose number operators are each given by an equal mixture of those for the two sublattices. Thus, the SBMFT constraints for each sublattice actually fixes the expectation value of the total number of $\alpha_\sigma$ and $\beta_\sigma$ modes. Therefore, a scattering process which creates $\alpha$ spinons and annihilates $\beta$ spinons (or vice versa) does not violate this constraint. 

For the AF on the simple cubic lattice $J_s^{\mathcal{A}} \neq J_s^{\mathcal{B}}$, and we get
\begin{multline}
	\label{J_s_a}
	J_s^{\mathcal{A}} = \frac{g_{\uparrow \downarrow} }{2S N^2}  \sum_{\boldsymbol{k}, \boldsymbol{k'}} \left\{ \sum_\nu \epsilon^{\nu \nu}_{\boldsymbol{k}\boldsymbol{k'}\uparrow \downarrow} \left[ n_1(\epsilon^{\nu}_{\boldsymbol{k}\uparrow}) - n_1(\epsilon^{\nu}_{\boldsymbol{k'}\downarrow}) \right] \left[ n_1(\nu \epsilon^{\nu \nu}_{\boldsymbol{k}\boldsymbol{k'}\uparrow \downarrow} ) - n_2(\nu \epsilon^{\nu \nu}_{\boldsymbol{k}\boldsymbol{k'}\uparrow \downarrow}) \right] \left( \delta_{\nu,+} u_{\boldsymbol{k} \uparrow}^2 u_{\boldsymbol{k'}\downarrow}^2 + \delta_{\nu,-} v_{\boldsymbol{k}\uparrow}^2 v_{\boldsymbol{k'}\downarrow}^2 \right) - \right. \\ \left. 
	\sum_\sigma \widetilde{\epsilon}^{+-}_{\boldsymbol{k}\boldsymbol{k'}\sigma \sigma} \left[ n_1(\epsilon^{+}_{\boldsymbol{k}\sigma}) + n_1(\epsilon^{-}_{\boldsymbol{k'}\sigma}) + 1\right] \left[ n_1(\widetilde{\epsilon}^{+-}_{\boldsymbol{k}\boldsymbol{k'}\sigma \sigma} ) - n_2(\widetilde{\epsilon}^{+-}_{\boldsymbol{k}\boldsymbol{k'}\sigma \sigma}) \right] \left( \delta_{\sigma, \uparrow} u_{\boldsymbol{k}\sigma}^2 v_{\boldsymbol{k'}\overline{\sigma}}^2 - \delta_{\sigma, \downarrow} u_{\boldsymbol{k'}\sigma}^2 v_{\boldsymbol{k}\overline{\sigma}}^2\right) \right\},
\end{multline}
where $\widetilde{\epsilon}^{+-}_{\boldsymbol{k}\boldsymbol{k'}\uparrow \uparrow} \equiv \epsilon^+_{\boldsymbol{k}\uparrow} + \epsilon^-_{\boldsymbol{k'}\uparrow}$, $\widetilde{\epsilon}^{+-}_{\boldsymbol{k}\boldsymbol{k'}\downarrow \downarrow} \equiv \epsilon^+_{\boldsymbol{k'}\downarrow} + \epsilon^-_{\boldsymbol{k}\downarrow}$, and $\overline{\nu} = -\nu$, the sum over $\nu$ runs over pseudospin indices, and the sum over $\sigma$ runs over spin indices. Similarly, we get
\begin{multline}
	\label{J_s_b}
	J_s^{\mathcal{B}} = \frac{g_{\uparrow \downarrow} }{2SN^2} \sum_{\boldsymbol{k}, \boldsymbol{k'}} \left\{ \sum_\nu \epsilon^{\nu \nu}_{\boldsymbol{k}\boldsymbol{k'}\uparrow \downarrow} \left[ n_1(\epsilon^{\nu}_{\boldsymbol{k}\uparrow}) - n_1(\epsilon^{\nu}_{\boldsymbol{k'}\downarrow}) \right] \left[ n_1(\nu \epsilon^{\nu \nu}_{\boldsymbol{k}\boldsymbol{k'}\uparrow \downarrow} ) - n_2(\nu \epsilon^{\nu \nu}_{\boldsymbol{k}\boldsymbol{k'}\uparrow \downarrow}) \right] \left( \delta_{\nu,-}  u_{\boldsymbol{k}\uparrow}^2 u_{\boldsymbol{k'}\downarrow}^2 + \delta_{\nu,+}v_{\boldsymbol{k}\uparrow}^2 v_{\boldsymbol{k'}\downarrow}^2 \right) - \right. \\ \left. 
	\sum_\sigma \widetilde{\epsilon}^{+-}_{\boldsymbol{k}\boldsymbol{k'}\sigma \sigma} \left[ n_1(\epsilon^{+}_{\boldsymbol{k}\sigma}) + n_1(\epsilon^{-}_{\boldsymbol{k'}\sigma}) + 1\right] \left[ n_1(\widetilde{\epsilon}^{+-}_{\boldsymbol{k}\boldsymbol{k'}\sigma \sigma} ) - n_2(\widetilde{\epsilon}^{+-}_{\boldsymbol{k}\boldsymbol{k'}\sigma \sigma}) \right] \left( \delta_{\sigma, \uparrow} u_{\boldsymbol{k'}\overline{\sigma}}^2 v_{\boldsymbol{k}\sigma}^2 - \delta_{\sigma, \downarrow} u_{\boldsymbol{k}\overline{\sigma}}^2 v_{\boldsymbol{k'}\sigma}^2 \right) \right\}.
\end{multline}
Note that the terms in the second lines of Eqs.~\eqref{J_s_a} and \eqref{J_s_b}, once combined in the sum $J_s \equiv J_s^{\mathcal{A}} + J_s^{\mathcal{B}}$, vanish quadratically in $B$ so they are negligibly small in the limit $B \ll \sqrt{JK}/\hbar \gamma$ (the spin-flop field). Thus, to linear order in $B$ we have
\begin{equation}
	J_s = \frac{g_{\uparrow \downarrow} }{2S N^2}  \sum_{\nu,\boldsymbol{k}, \boldsymbol{k'}}\epsilon^{\nu \nu}_{\boldsymbol{k}\boldsymbol{k'}\uparrow \downarrow} \left[ n_1(\epsilon^{\nu}_{\boldsymbol{k}\uparrow}) - n_1(\epsilon^{\nu}_{\boldsymbol{k'}\downarrow}) \right] \left[ n_1(\nu \epsilon^{\nu \nu}_{ \boldsymbol{k}\boldsymbol{k'}\uparrow \downarrow} ) - n_2(\nu \epsilon^{\nu \nu}_{\boldsymbol{k}\boldsymbol{k'}\uparrow \downarrow}) \right] ( u_{\boldsymbol{k} \uparrow}^2 u_{\boldsymbol{k'}\downarrow}^2 + v_{\boldsymbol{k}\uparrow}^2 v_{\boldsymbol{k'}\downarrow}^2 ).
\end{equation}
The condensate contribution to the spin current in linear response reproduces the equation for $\mathcal{S}^{\mathrm{AF}}$ in the main text. It comes from separating $n^+_{\uparrow}$ ($n^-_{\downarrow}$) at the points in the BZ where $\epsilon^+_{\uparrow} = 0$ ($\epsilon^-_{\downarrow} = 0$).

\textit{Holstein-Primakoff spin currents.}| The Holstein-Primakoff approximation (defined as the Holstein-Primakoff transformation expanded to second order in the magnon over spin densities) results are obtained analogously to the SBMFT. The HPA spin current is $J_s = J_s^{\mathcal{A}} + J_s^{\mathcal{B}}$, where $J_s^{\mathcal{A}}$ is generated by the coupling $H^{\mathcal{A}}_{\mathrm{int}} = (V/\sqrt{N}) \sum_{\boldsymbol{k},q,q'} \tilde{a}_{\boldsymbol{k}} c_{q \uparrow} c^{\dagger}_{q'\downarrow}+ \mathrm{H.c}$, and $J_s^{\mathcal{B}}$ by $H^{\mathcal{B}}_{\mathrm{int}} = (V/\sqrt{N}) \sum_{\boldsymbol{k},q,q'} \tilde{b}_{\boldsymbol{k}} c_{q \uparrow} c^{\dagger}_{q'\downarrow} + \mathrm{H.c}$, where $\tilde{a}, \tilde{b}$ are the magnon operators on the $\mathcal{A}, \mathcal{B}$ sublattices, respectively. These results are a special case of the more general Holstein-Primakoff results derived in Ref.\cite{Flebus_2019}, where the interfacial Hamiltonian is treated as the sum of interfacial exchange coupling to each sublattice with separate exchange constants. Our Holstein-Primakoff spin current can be obtained from their general result as follows: $J_s^{\mathcal{A}}$ is obtained by taking their exchange constant for the $\mathcal{B}$ lattice to be zero, and $J_s^{\mathcal{B}}$ by taking their exchange constant for the $\mathcal{A}$ lattice to be zero; then $J_s$ is the sum of these two incoherent contributions weighted by the same exchange constant.

\bibliography{sbmft}